\documentclass[preprint,floats,aps,epsfig,nofootinbib,amssymb]{revtex4}
\usepackage{graphicx}
\usepackage{dcolumn}
\usepackage{bm}
\usepackage{float}
\usepackage{mathrsfs}
\usepackage{slashed}
\usepackage{graphicx,color}
\usepackage{epsfig}
\usepackage{subfigure}
\usepackage{color}
\usepackage{amsmath}
\usepackage{amssymb}
\usepackage{caption2}
\usepackage{amsmath}
\usepackage{indentfirst}
\usepackage{amsmath}
\begin{document}

\title{Looking for New Physics via Semi-leptonic and Leptonic rare decays of $D$ and $D_s$}
\author{Xing-Dao Guo$^{1,4}$}
\email{guoxingdao@mail.nankai.edu.cn}
\author{Xi-Qing Hao$^2$}
\email{haoxiqing@htu.edu.cn}
\author{Hong-Wei Ke$^3$}
\email{khw020056@tju.edu.cn}
\author{Ming-Gang Zhao$^4$}
\email{zhaomg@nankai.edu.cn}
\author{Xue-Qian Li$^4$}
\email{lixq@nankai.edu.cn}
\author{}
\author{}
\author{}
\affiliation{1. School of Physics and Math, Xuzhou University of Technology, Xuzhou, 221111, P.R. China\\
2. Physics Department, Henan Normal University, Xinxiang 453007, P.R. China\\
3. School of Science, Tianjin University, Tianjin, 300072, P.R. China \\
4. Department of Physics, Nankai University, Tianjin, 300071, P.R. China}

\begin{abstract}

It is well recognized that looking for new physics at lower energy
colliders is a tendency which is complementary to high energy
machines such as LHC. Based on large database of BESIII, we may have
a unique opportunity to do a good job. In this paper we calculate
the branching ratios of  semi-leptonic processes $D^+_s \to K^+
e^-e^+$, $D^+_s \to K^+ e^-\mu^+$ and leptonic processes $D^0 \to
e^-e^+$, $D^0 \to e^-\mu^+$ in the frames of $U(1)'$ model, 2HDM and
unparticle separately. It is found that both the $U(1)'$ and 2HDM may influence the semi-leptonic
decay rates, but only the $U(1)'$ offers
substantial contributions to the pure leptonic decays and the resultant branching ratio of $D^0
\to e^-\mu^+$ can be as large as  $10^{-7}\sim10^{-8}$ which might
be  observed at the future super $\tau$-charm factory.

\end{abstract}

\maketitle

\section{Introduction}

One of tasks of the colliders with high-intensity but lower-energy is to find
traces of new physics beyond the Standard Model(SM) through
measuring the rare decays with high accuracy, namely look for
deviations of  the measured values from the SM predictions.
Generally, it is believed that new physics scale may exist at
several hundreds of GeV to a few TeV whereas for lower energies, the
contributions from new physics might be drowned out in the SM background. However, in
some rare decays, contributions from SM are highly suppressed or
even forbidden, then the  new physics beyond SM (BSM) might emerge and play the
leading role. If such processes are observed in high precision
experiments, a trace of BSM could be pinned down. Concretely,  the
processes where the flavor-changing-neutral-current (FCNC) is
involved, are the goal of our studies. Even though such results may
not determine what kind of new physics, it may offer valuable
information about new physics to the high energy colliders such as
LHC. In SM, FCNC and lepton flavor violation(LFV) processes can only
occur via loop diagrams so would suffer a suppression. Thus study on
the FCNC/LFV transitions would compose a key for the BSM search.

The rare decays of D and B mesons provide a favorable area because
they are produced at $e^+e^-$ colliders, where the background is
much cleaner than that at hadron colliders. The newest
measurements set upper bounds for the branching ratios of $D^+_s \to
K^+ e^-e^+$ and $D^+_s \to K^+ e^-\mu^+$ as $3.7\times10^{-6}$ and
$9.7\times10^{-6}$ respectively \cite{Olive:2016xmw}, and the upper
bounds for $D^0 \to e^-e^+$ and $D^0 \to e^-\mu^+$ are
$7.9\times10^{-8}$ and $2.6\times10^{-7}$ \cite{Olive:2016xmw}.
Theoretically, those decay processes receive contributions from both
short and long distance effects of SM \cite{Burdman:2001tf}. Especially,
for  $D^+_s \to K^+ e^-e^+$, its rate mainly is determined by the
long distance effect and the SM predicted value is
$1.8\times10^{-6}$, which is higher than the short distance
contribution ( $2\times10^{-8}$\cite{Burdman:2001tf}) by two orders.
For other concerned processes, the contributions from SM are so
small that can be neglected.

As indicated, at lower energy experiments, one can notice the new
physics trace, but cannot determine what it is, thus in
collaboration, theorists would offer possible scheme(s) to
experimentalists and help them to extract information from the data.
That is the main idea of this work.

There are many new physics models (BSM) constructed by numerous
theorists, for example, the fourth generation\cite{Hou:2006mx}, the
non-universal $Z'$
boson\cite{Langacker:2008yv,Leike:1998wr,rnm,Yue:2016mqm}, the 2
Higgs doublet
model(2HDM)\cite{Cheng:1987rs,Davidson:2010xv,Omura:2015nja} and the
unparticle\cite{Georgi:2007ek,Georgi:2007si,Luo:2007bq} etc., in
their framework, FCNC/LFV processes occur at tree level. Thus once
such rare decays involving FCNC/LFV processes are experimentally
observed, one may claim existence of BSM, then comparing the values
predicted by different models with the data, he would gain a hint
about what BSM may play role which is valuable  for high
energy colliders.

In Refs.\cite{Fajfer:2015mia,deBoer:2015boa}, based on several BSM
models the authors derived the formulaes and evaluated the decay
rates of semi-leptonic and leptonic decays of $D$ mesons while the
model parameters are constrained mainly by the data of $D^0-\bar D^0$
mixing. The result obtained by them was pessimistic that  these
decay rates cannot provide any trace of the concerned models. In
this work we choose three new physics models: $U(1)'$ model, 2HDM
type III and unparticle but relax the constraint from $D^0-\bar D^0$
mixing by supposing there were some unknown reasons to suppress the
rate if the present measurements are sufficiently accurate, instead
we consider the constraints obtained by fitting the experimental
data for $\tau\to 3l$\cite{Olive:2016xmw,Yue:2016mqm}. Then we calculate the branching ratios of $D^+_s \to K^+
e^-e^+$, $D^+_s \to K^+ e^-\mu^+$, $D^0 \to e^-e^+$ and $D^0 \to
e^-\mu^+$ in the framework of those models respectively.  Our
numerical results show that only $Z'$ which is from a broken extra
$U'(1)$ gauge symmetry and 2HDM of type III can result in substantial enhancement to the
branching ratios of $D^+_s \to K^+ e^-e^+$ and $D^+_s \to K^+
e^-\mu^+$  up to $10^{-6}\sim10^{-7}$. Those results will be tested
in future BES III experiment. Indeed ,we lay our hope on the huge
database of BES III, without which we cannot go any further to search
for new physics after all.

We, in this work, also try to set schemes for analyzing the data on
those decays based on the BES III data and extract information about new
physics BSM.

This paper is organized as follows. In Sections 2 and 3, we first
briefly review the SM results for the semi-leptonic and pure leptonic rare decays
and then derive corresponding contributions induced by new physics models: extra
$U(1)'$ , 2HDM of type III and unparticle one by one. In fact some of them had been deduced
by other authors and here we only probe their formulation, moreover add those which were not derived
before. We obtain the
corresponding Feynman amplitudes and decay widths for $D^+_s \to K^+
e^-e^+$, $D^+_s \to K^+ e^-\mu^+$, $D^0 \to e^-e^+$ and $D^0 \to
e^-\mu^+$. In section 4, we present our numerical results along with the
constraints on the model parameters obtained by fitting previous
experimental data except the $D^0-\bar D^0$ mixing. In section 5, we
set an experimental scheme for analyzing the data which will be
achieved by the BES III collaborations in the near future. In the last section, we
present a brief discussion and draw our conclusion.

\bibliographystyle{prsty}

\section{$D^+_s$ semi-leptonic decay}

For the decay processes $D^+_s \to K^+ e^-e^+$ and $D^+_s \to K^+
e^-\mu^+$, the contributions of SM  to these  FCNC  processes are
realized via electromagnetic penguin diagrams and suppressed.
However, besides the short-distance effects, there exist a
long-distance contribution which is larger. Moreover, because of
smallness of the direct SM process, any new physics model whose
Hamiltinian includes  FCNC interactions, may induce  the
semi-leptonic and leptonic decays of $D^+_s$ and $D^0$ at tree
level. In this section we only explore three possible models:
$U(1)'$ model, 2HDM of type III and unparticle. Since those models have
been studied by many authors from various aspects, here we only give
a brief review.

\subsection{the SM contribution}

\begin{figure}[H]
\centering
\begin{minipage}[t]{0.39\linewidth}
\centering
\includegraphics[width=1.0\textwidth]{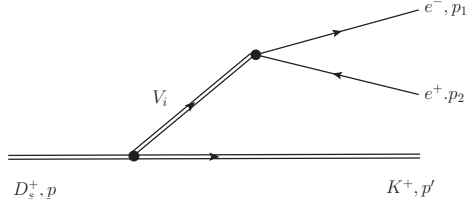}
\end{minipage}
\caption{The Feynman diagrams of process $D^+_s \to K^+ e^-e^+$
through SM long distance. } \label{eesm}
\end{figure}

The authors of
Refs.\cite{Burdman:2001tf,deBoer:2015boa,Buras:1998raa} gave the
amplitudes for $D^+_s \to K^+ e^-e^+$, here we only list the
formulas for readers' convenience. The Feynman amplitude of decay
$D^+_s \to K^+ e^-e^+$ in the framework of SM is
\begin{equation}
\begin{array}{rl}
\mathcal{M}_{SM}=&\frac{4G_F}{\sqrt 2}[C_7\langle e^+e^- |e A^\delta \bar l\gamma_\delta l|\gamma\rangle\frac{1}{q^2}
\langle \gamma K^+|O_7|D^+_s\rangle+
C_9\langle e^+e^- K^+|O_9|D^+_s\rangle]\\
\end{array}
\end{equation}
where
\begin{equation}
\begin{array}{rl}
O_7=&\frac{e}{16\pi^2}m_c(\bar u_L \sigma^{\alpha\beta}c_R)F_{\alpha\beta}\\
O_9=&\frac{e^2}{16\pi^2}(\bar u_L \gamma^\alpha c_L)\bar l\gamma_\alpha l \\
\end{array}
\end{equation}
After some simple reductions, $\mathcal{M}_{SM}$ is transited to
\begin{equation}
\begin{array}{rl}
\mathcal{M}_{SM}=&\frac{4G_F}{\sqrt 2}\frac{e^2m_c}{16\pi^2}C_7\frac{\bar u(p_2)(\gamma_\beta q_\alpha-\gamma_\alpha q_\beta)v(p_1)}{2q^2}
\frac{f_T(q^2)}{m_{D_s}}[(p+p')^\alpha q^\beta-(p+p')^\beta q^\alpha+i\epsilon^{\alpha\beta\rho\sigma}(p+p')_\rho q_\sigma]\\
&+\frac{e^2}{32\pi^2}C_9\bar u(p_2)\gamma^\delta v(p_1)\{f_+(q^2)[(p+p')_\delta-\frac{m^2_{D_s}-m^2_K}{q^2}q_\delta]+f_0(q^2)\frac{m^2_{D_s}-m^2_K}{q^2}q_\delta\}
\end{array}
\end{equation}
where $q=p_1+p_2$, $C_7=4.7\times10^{-3}$\cite{Khodjamirian:2009ys}.
Following Refs.\cite{Burdman:2001tf,deBoer:2015boa}, we also
consider the resonance processes $D^+_s \to K^+ V_i\to K^+ e^-e^+$
with $i=\rho,\omega,\phi$ which are accounted as long distance
contributions and the corresponding Feynman diagrams are shown in
Fig.\ref{eesm}.

Thus $C_9$ can be written as
\begin{equation}
C_9=(0.012+\frac{3\pi}{\alpha_e^2}\sum_{i=\rho,\omega,\phi}\kappa_i\frac{m_{V_i}\Gamma_{V_i\to e^+e^-}}{m_{V_i}^2-q^2-im_{V_i}\Gamma_{V_i}})
(V_{ud}V_{cd}+V_{us}V_{cs})
\end{equation}
with $\kappa_\rho=0.7$, $\kappa_\omega=3.1$ and $\kappa_\phi=3.6$.
The second part in the parenthesis corresponds to the long-distance
contributions.

Following Ref.\cite{Fajfer:2015mia,Khodjamirian:2009ys}, the
hadronic form factors are written as
\begin{equation}
\begin{array}{rl}
&f_T(q^2)=\frac{f^T_{D_s K}(0)}{(1-q^2/m^2_{D_s})(1-a_T q^2/m^2_{D_s})} \\
&f_+(q^2)=\frac{f^+_{D_s K}(0)}{(1-q^2/m^2_{D_s})(1-\alpha_{D_s K}q^2/m^2_{D_s})} \\
&f_0(q^2)=\frac{f^+_{D_s K}(0)}{1-q^2/(\beta_{D_s K}m^2_{D_s})} \\
\end{array}
\end{equation}
where $f^T_{D_s K}(0)=0.46$, $a_T=0.18$, $f^+_{D_s K}(0)=0.75\pm0.08$, $\alpha_{D_s K}=0.30\pm0.03$ and $\beta_{D_s K}=1.3\pm0.07$.

The long-distance contribution is of an order of $10^{-6}$
\cite{Burdman:2001tf}. Thus the contribution from SM may be close or
even larger than that of BSM, so they would interfere among each
other. We will discuss it in section 4.

\subsection{Contributions of $Z'$ in the $U(1)'$ model}

The $U(1)'$ model was proposed and applied by many authors
\cite{Gao:2010zzg,Langacker:2008yv,Leike:1998wr,Langacker:2009im},
and the corresponding lagrangian is
\begin{equation}
\mathcal{L}_{Z'}=\sum_{i,j}[\bar l_i \gamma^\mu(\omega_{ij}^L P_L+\omega_{ij}^R P_R) l_j Z'_\mu +
\bar q_i\gamma^\mu(\varepsilon_{ij}^L P_L+\varepsilon_{ij}^R P_R) q_j Z'_\mu]+h.c.
\end{equation}
where $P_{L(R)}=\frac{1-(+)\gamma_5}{2}$, $\omega_{ij}$ (
$\varepsilon_{ij}$) denote the chiral couplings between the new
gauge boson $Z'$ and various leptons (quarks). Whether  it can
be applied to solve some phenomenological anomalies, the key point is the  intensity of the
coupling and the mass of $Z'$ gauge boson which would be fixed by fitting available data.

\begin{figure}[H]
\centering
\begin{minipage}[t]{0.39\linewidth}
\centering
\includegraphics[width=1.0\textwidth]{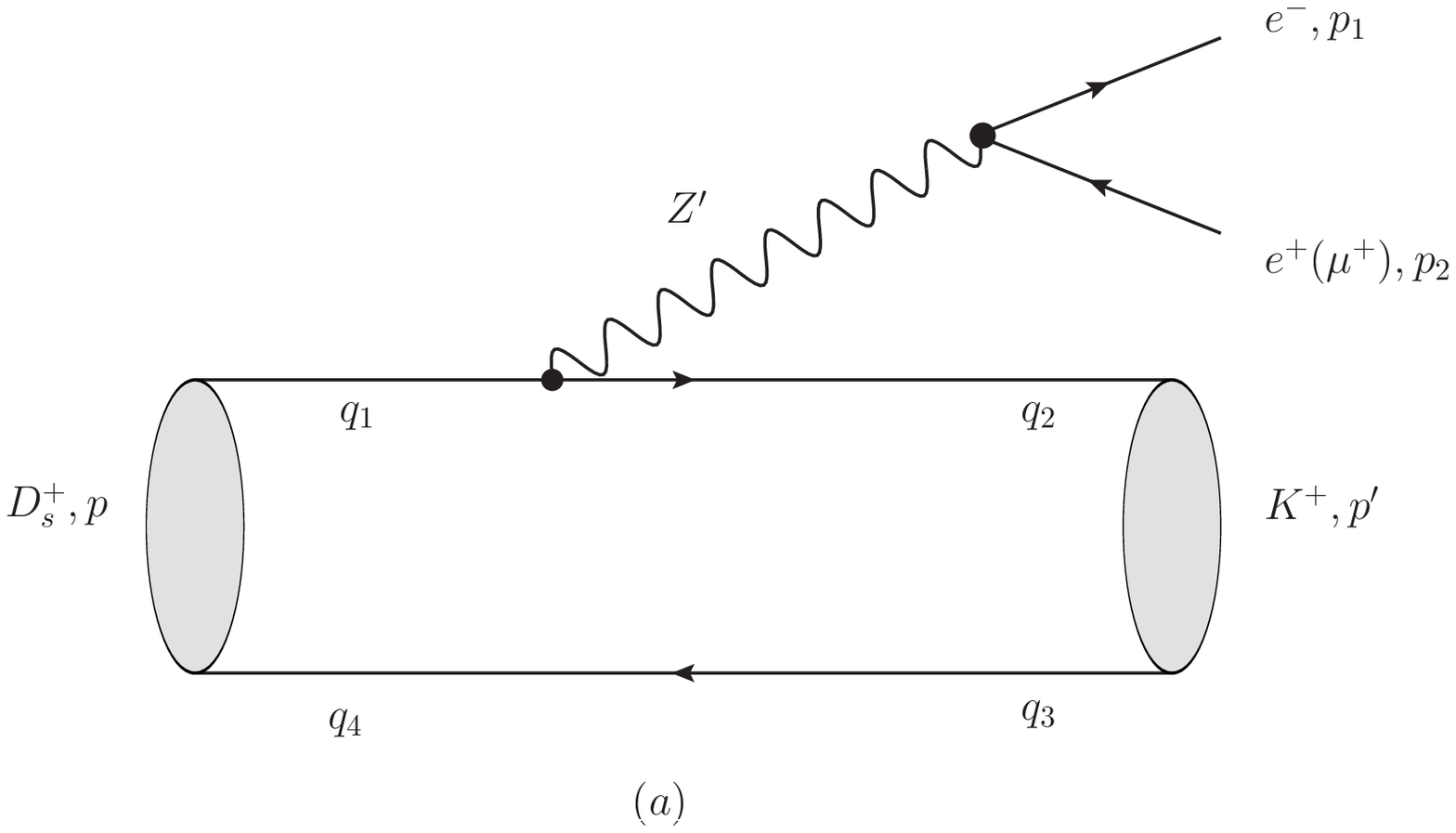}
\end{minipage}
\begin{minipage}[t]{0.39\linewidth}
\centering
\includegraphics[width=1.0\textwidth]{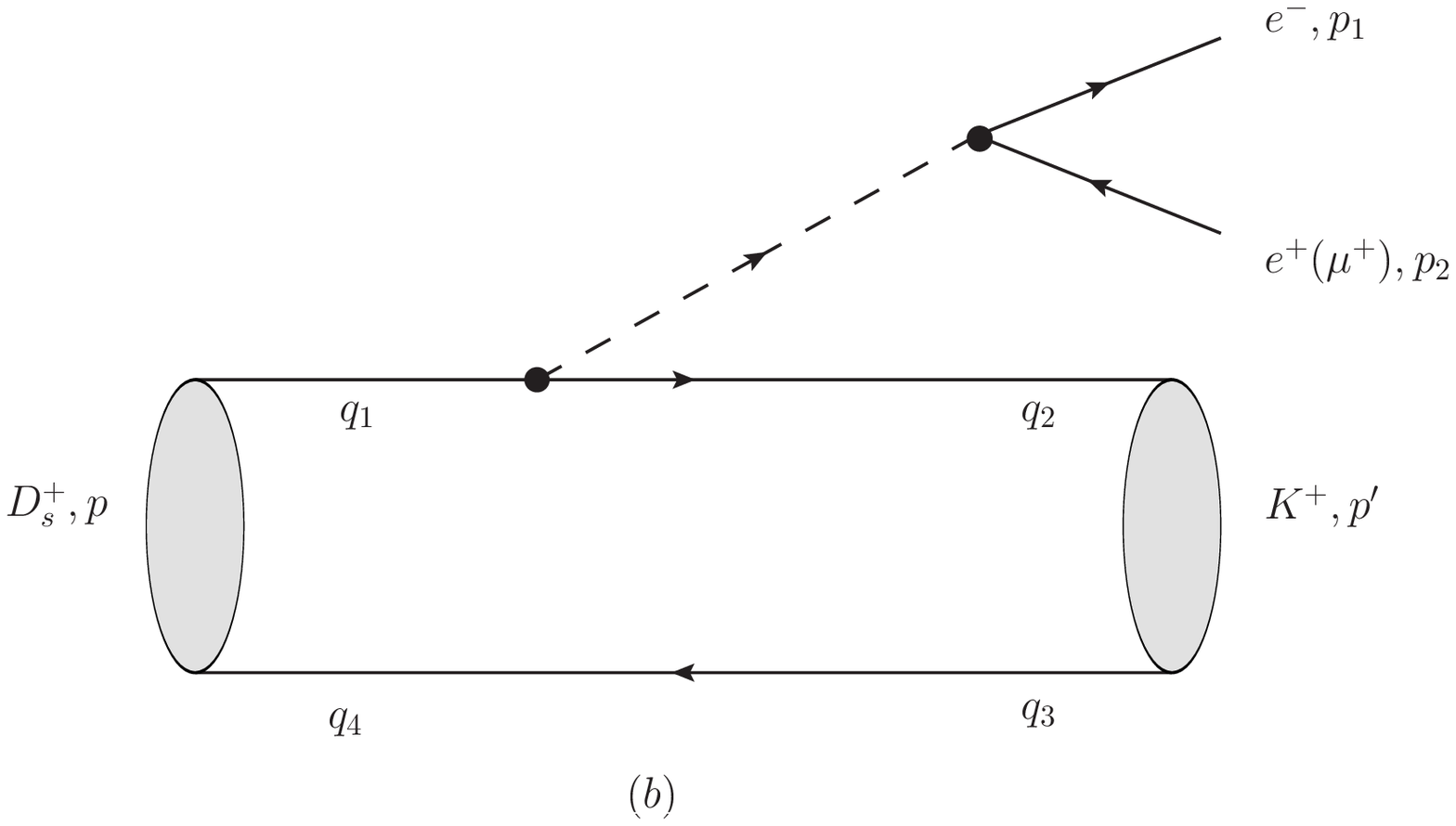}
\end{minipage}
\begin{minipage}[t]{0.39\linewidth}
\centering
\includegraphics[width=1.0\textwidth]{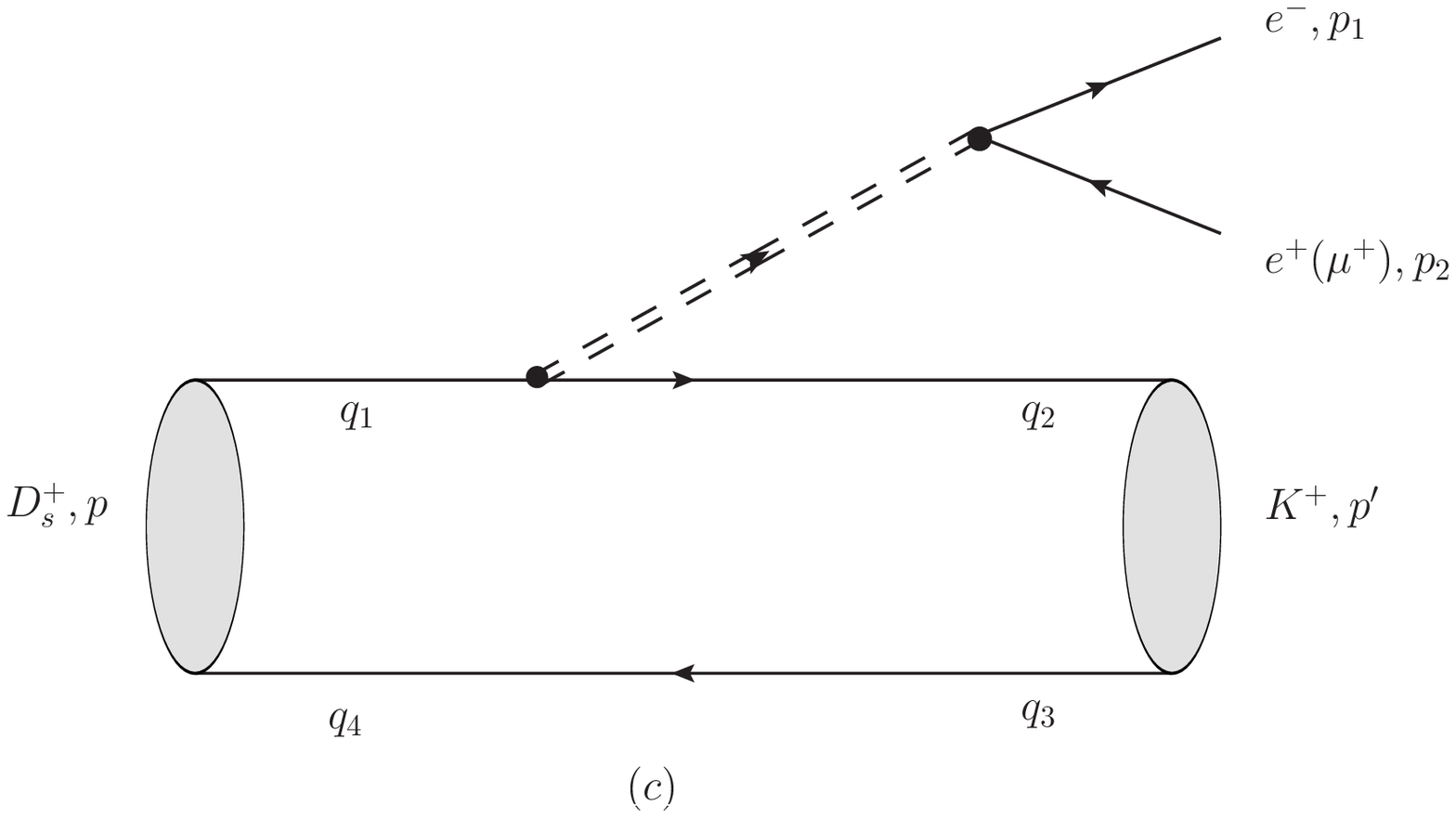}
\end{minipage}
\caption{The Feynman diagrams for $D^+_s \to K^+ e^-e^+$ and $D^+_s
\to K^+ e^-\mu^+$ in $U(1)'$ model (a), 2HDM type III (b) and
unparticle (c) respectively.} \label{eehh}
\end{figure}

For the decay processes $D^+_s \to K^+ e^-e^+$ and $D^+_s \to K^+
e^-\mu^+$, corresponding Feynman diagrams are shown in
Fig.\ref{eehh}.

The corresponding Feynman amplitude with $Z'$ as the mediate
particle was derived by the authors of
\cite{Gao:2010zzg,Langacker:2008yv,Leike:1998wr,Langacker:2009im} as
\begin{equation}
\begin{array}{rl}
\mathcal{M}_{Z'}(D^+_s \to K^+ l_i \bar l_j)=
&\{f_+(q^2)[(p+p')_\sigma-\frac{m^2_{D_s}-m^2_K}{q^2}q_\sigma]+f_0(q^2)\frac{m^2_{D_s}-m^2_K}{q^2}q_\sigma\}\\
&\frac{\varepsilon_{cu}^L +\varepsilon_{cu}^R}{g/\sqrt 2}\frac{1}{q^2-m_{Z'}^2}
\bar u(p_2)(\omega_{ij}^L P_L+\omega_{ij}^R P_R)\gamma^\sigma v(p_1)
\end{array}
\end{equation}
where $\omega_{ij}=\omega_{ee}$ for  $D^+_s \to K^+ e^-e^+$ and
$\omega_{ij}=\omega_{e\mu}$ for  $D^+_s \to K^+ e^-\mu^+$
respectively.

The contributions of SM (indeed from the long-distance part) and
$Z'$ might be of the same order depending on the model parameters£¬
thus we should consider their interference. So we have
\begin{equation}
\begin{array}{rl}
|\mathcal{M}|^2=&|\mathcal{M}_{SM}+\mathcal{M}_{Z'} e^{i\phi}|^2\\
=&|\mathcal{M}_{SM}|^2+|\mathcal{M}_{Z'}|^2+2|\mathcal{M}_{SM}
\mathcal{M}_{Z'}| \cos\phi.
\end{array}
\end{equation}
Averaging initial spin and summing over finial spin polarizations,
the decay width $\Gamma(D^+_s \to K^+ e^-e^+)$ is
\begin{equation}
\begin{array}{rl}
\frac{d\Gamma}{dq^2}=&[\frac{G^2_F \alpha^2_e}{1536\pi^5 m^3_{D_s}}|C_9 f_+(q^2))^2+2C_7 f_T(q^2) \frac{m_c}{m_{D_s}}|^2
+\frac{(\varepsilon_{cu}^L+\varepsilon_{cu}^R)^2((\omega_{ee}^L)^2+(\omega_{ee}^R)^2)}{192 \pi^3 g^2 m^4_{Z'} m^3_{D_s}}f_+(q^2)^2\\
&+\frac{(\varepsilon_{cu}^L+\varepsilon_{cu}^R)(\omega_{ee}^L+\omega_{ee}^R)G_F \alpha_e}{384\pi^4 g m^2_{Z'} m^3_{D_s}}
f_+(q^2)(C_9 f_+(q^2))^2+2C_7 f_T(q^2) \frac{m_c}{m_{D_s}})\cos\phi]
\lambda^{3/2}(q^2,m^2_{D_s},m^2_K)
\end{array}
\end{equation}
where $\lambda(a,b,c)=a^2+b^2+c^2-2ab-2bc-2ca$ is the Kallen
function. We can obtain the total decay width by integrating over
$dq^2$, as
\begin{equation}
\begin{array}{rl}
\Gamma=\int^{(m_{D_s}-m_K)^2}_{4m_e^2}\frac{d\Gamma}{dq^2} dq^2
\end{array}
\label{tot}
\end{equation}

\subsection{Contributions of heavy neutral Higgs in the two-Higgs-Doublet Model of type III}
In 2HDM of type III
\cite{Davidson:2010xv,Omura:2015nja,Liu:2015oaa}, there are two
neutral CP even Higgs bosons, one is the Higgs boson in SM and
another is a heavy Higgs boson, the corresponding Lagrangian for the
heavy Higgs boson is
\begin{equation}
\mathcal{L}_{Yukawa}=\sum_{i,j}[\bar l_i (\frac{m^i_l}{v}\cos\alpha \delta_{ij}-\frac{\rho^E_{ij}}{\sqrt 2}\sin\alpha) l_j H +
 \bar q_i(\frac{m^i_q}{v}\cos\alpha \delta_{ij}- \frac{\rho^U_{ij}}{\sqrt 2}\sin\alpha) q_j H]+h.c.
\end{equation}
where $\rho^E_{ij}$ and $\rho^U_{ij}$ stand for effective coupling
constants for leptons and quarks respectively. $\cos\alpha$ is the
mixing angle between
light and heavy Higgs bosons. Following Refs. \cite{Omura:2015nja,Liu:2015oaa}, we take $\cos\alpha \to 0$.
and do not adopt the  so-called Cheng¨CSher ansatz for $\rho^f_{ij}$
which was discussed in Ref.\cite{Cheng:1987rs}. Instead,  we take a range of
$\rho^f_{ij}$ to $0.1\sim0.3$ as suggested in Ref. \cite{Liu:2015oaa}.

The Feynman amplitude corresponding to contributions  through
exchanging a heavy Higgs boson is
\begin{equation}
\begin{array}{rl}
\mathcal{M}_{hh}(D^+_s \to K^+ l_i \bar l_j)
=&\{2f^+_{D_s K}(q^2)\frac{p'\cdot p}{m_{D_s}}+[f^+_{D_s K}(q^2)+f^-_{D_s K}(q^2)
]\frac{q\cdot p}{m_{D_s}}\}\\
&\rho^U_{cu}\frac{1}{q^2-m^2_{hh}} \bar u(p_1)v(p_2)\rho^E_{ij}
\end{array}
\end{equation}
where $\rho_{ij}=\rho_{ee}$, $\rho_{ij}=\rho_{e\mu}$  stand for
$D^+_s \to K^+ e^-e^+$ and $D^+_s \to K^+ e^-\mu^+$ respectively.

The differential decay width $d\Gamma(D^+_s \to K^+ e^-e^+)$ is
\begin{equation}
\begin{array}{rl}
\frac{d\Gamma}{d q^2}=&[\frac{G^2_F \alpha^2_e}{1536\pi^5 m^3_{D_s}}|C_9 f_+(q^2))^2+2C_7 f_T(q^2) \frac{m_c}{m_{D_s}}|^2\lambda(q^2,m^2_{D_s},m^2_K)\\
&+\frac{(\rho^U_{cu}\rho^E_{ee})^2 (f^0_{D_s K}(q^2)(m^2_{D_s}-m^2_K)(m^2_{D_s}-m^2_K+s_{12})
-f^+_{D_s K}(q^2)(m_{D_s}^4+(m_K^2-s_{12})^2-2m_{D_s}^2(m_K^2+s_{12})))^2}{64 g^2 m_{D_s}^5 m_{hh}^4 \pi^3 s12}]\\
&\lambda^{1/2}(q^2,m^2_{D_s},m^2_K).
\end{array}
\end{equation}
Then we obtain the total decay width by integrating over $dq^2$ as
done in Eqn.\ref{tot}.

\subsection{contribution from unparticle}

The idea of unparticle was proposed by Georgi\cite{Georgi:2007ek} a
while ago. Then many authors followed him to explore relevant
phenomenology and study the basic theory. In the scenario of unparticle,
flavor changing term exists in the basic Lagrangian, so that the
FCNC can occur at tree level. One is naturally tempted to conjecture
that the unparticle mechanism may contribute to $D^+_s \to K^+
e^-e^+$ and $D^+_s \to K^+ e^-\mu^+$. Following
Ref.\cite{Georgi:2007si,Li:2007by,Jia:2013haa}, we only consider the
interactions between fermions and scalar unparticle. The
corresponding effective interaction is :
\begin{equation}
\mathcal{L}=\sum_{f',f}\frac{c^{f'f}_s}{\Lambda_\mathcal{U}^{d_\mathcal{U}}}\bar f'\gamma_\mu (1-\gamma_5)f \partial^\mu \mathcal{O}_\mathcal{U}+h.c.
\end{equation}
where $c^{f'f}_s$ stands for coupling constants between unparticle
and fermions, $\mathcal{O}_\mathcal{U}$ is the scalar unparticle
field, $d_\mathcal{U}$ is a nontrivial scale dimension and
$\Lambda_\mathcal{U}$ is an energy scale at order of TeV. The
propagator of the scalar unparticle
is\cite{Jia:2013haa,Cheung:2007zza,Luo:2007bq}
\begin{equation}
\begin{array}{rl}
\int d^4x e^{iP\cdot x} \langle 0|T \mathcal{O}_\mathcal{U}(x)
\mathcal{O}_\mathcal{U}(0)|0\rangle=
i\frac{A_{d_\mathcal{U}}}{2\sin(d_\mathcal{U}\pi)}(P^2)^{2-d_\mathcal{U}}e^{-i(d_\mathcal{U}-2)\pi},
\end{array}
\end{equation}
with $A_{d_\mathcal{U}}$ is
\begin{equation}
\begin{array}{rl}
A_{d_\mathcal{U}}=\frac{16\pi^{5/2}}{(2\pi)^{2d_\mathcal{U}}} \frac{\Gamma(d_\mathcal{U}+1/2)}{\Gamma(d_\mathcal{U}-1) \Gamma(2d_\mathcal{U})}.
\end{array}
\end{equation}

Supposing $D^+_s \to K^+ e^-e^+$ and $D^+_s \to K^+ e^-\mu^+$ occur
via exchanging a scalar unparticle, the corresponding Feynman
amplitude is
\begin{equation}
\begin{array}{rl}
\mathcal{M}(D^+_s \to K^+ l_i \bar l_j)=&\{2f^+_{D_s K}(q^2)p'\cdot q+[f^+_{D_s K}(q^2)+f^-_{D_s K}(q^2)]q^2\}\\
&\frac{c^{cu}_s}{\Lambda_\mathcal{U}^{d_\mathcal{U}}}\frac{1}{(q^2)^{2-d_\mathcal{U}}}e^{-i(d_\mathcal{U}-2)\pi}
\bar u(p_1)\rlap /q
(1-\gamma_5)v(p_2)\frac{c^{ij}_s}{\Lambda_\mathcal{U}^{d_\mathcal{U}}}.
\end{array}
\end{equation}
where $c^{ij}_s=c^{ee}_s$, $c^{ij}_s=c^{e\mu}_s$ correspond to
$D^+_s \to K^+ e^-e^+$ and $D^+_s \to K^+ e^-\mu^+$ respectively.

Since numerically the unparticle contribution to $D^+_s \to K^+
e^-e^+$ and $D^+_s \to K^+ e^-\mu^+$ is much smaller than that from
SM and other models BSM, we list the formula involving
unparticle, and for completeness, we include the numerical results of the unparticle
contribution in the corresponding tables. The differential
decay width $\Gamma(D^+_s \to K^+ e^-e^+)$ is
\begin{equation}
\begin{array}{rl}
\frac{d\Gamma}{d q^2}=&\frac{1}{256\pi^3 m^3_{D_s}}(c^{cu}_s c^{ee}_s)^2
\frac{2^{12-4d_\mathcal{U}}me^2\pi^{5-4d_\mathcal{U}}(2me^2+s_{12})}{s_{12}^{6-2d_\mathcal{U}}}
\frac{\Gamma^2[1/2+d_\mathcal{U}]}{\Lambda_\mathcal{U}^{4d_\mathcal{U}}\sin^2 d_\mathcal{U}\pi}\\
&\frac{(f^0_{D_s K}(q^2)(m^2_{D_s}-m^2_K)(2m_e^2+s_{12})+2f^+_{D_s
K}(q^2)m_e^2(-m^2_{D_s}+m^2_K+s_{12}))^2}
{g^2\Gamma^2[d_\mathcal{U}-1]\Gamma^2[2d_\mathcal{U}]}\lambda^{1/2}(q^2,m^2_{D_s},m^2_K).
\end{array}
\end{equation}

\subsection{ Semi-leptonic decay of $D^+$}

Decays of $D^+ \to \pi^+ e^-e^+$ and $D^+ \to \pi^+ e^-\mu^+$ are
similar to  $D^+_s \to K^+ e^-e^+$ and $D^+_s \to K^+ e^-\mu^+$,
only difference is the species of the spectators. Therefore all the
formulas of $D^+_s \to K^+ l_i \bar l_j$ can be transferred to $D^+
\to K^+ l_i \bar l_j$ by an $SU(3)$ symmetry.

\section{ Rare Leptonic decays of $D^0$}

The rare leptonic decays of $D^0$ refer to $D^0\to l\bar l$ and $D^0\to
l_i\bar l_j$ with $i\neq j$ which is not only a FCNC, but also a lepton-flavor violation (LFV)
process.  In SM, in $D^0\to l\bar l$, charm-quark and $\bar u$
annihilate into a virtual photon via an electromagnetic penguin
which suppresses the reaction rate. For the LFV process, not only at
the initial part, $c$ and $\bar u$ need annihilating into a Z
virtual meson which later turns into a pair of neutrinos, then via a
weak scattering the neutrinos eventually end with two leptons with
different flavors. Because neutrinos are very light, this process is
much suppressed than $D^0\to l\bar l$. In fact, if there does not
exist new physics BSM, such LFV processes can never be
experimentally measured. Therefore, search for such LFV processes composes
a trustworthy probe of BSM.  Actually, contribution to the leptonic decays
(both lepton-flavor conserving and lepton-flavor violating
processes) of SM is too small to be observed\cite{Burdman:2001tf},
thus we only consider contribution from new physics. Since $D^0$ is a
pseudo-scalar meson and heavy Higgs is scalar boson, processes $D^0
\to e^-e^+$ and $D^0 \to e^-\mu^+$ cannot occur through exchanging
heavy Higgs boson. In the $Z'$ and unparticle scenarios  $D^0 \to e^-e^+$
and $D^0 \to e^-\mu^+$ might be induced to result in sizable rates.

\subsection{The $Z'$ gauge boson from $U(1)'$ model}

\begin{figure}[H]
\centering
\begin{minipage}[t]{0.45\linewidth}
\centering
\includegraphics[width=1.0\textwidth]{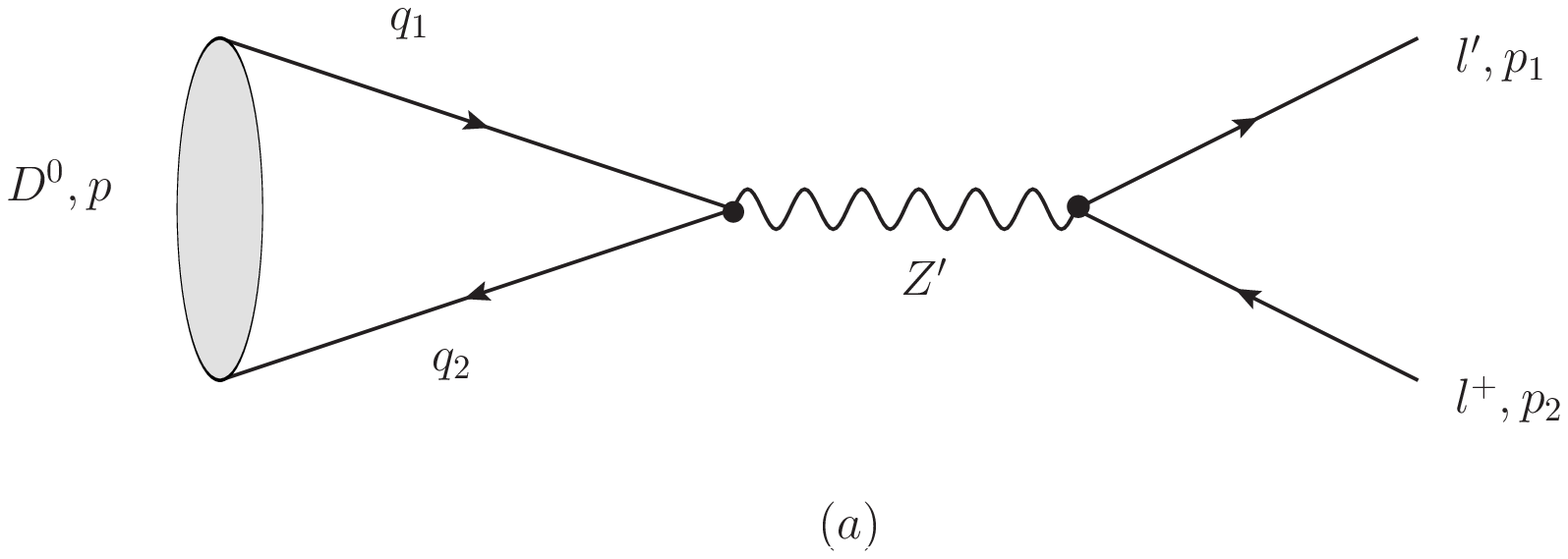}
\end{minipage}
\begin{minipage}[t]{0.45\linewidth}
\centering
\includegraphics[width=1.0\textwidth]{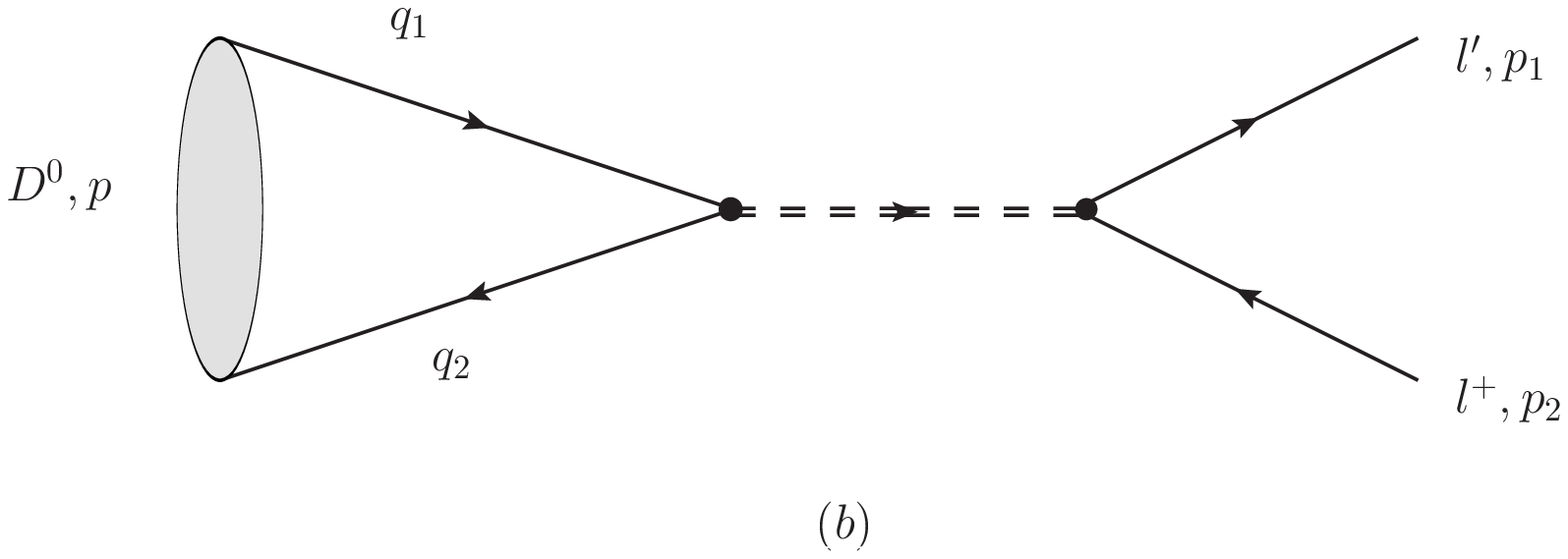}
\end{minipage}
\caption{The Feynman diagrams of processes $D^0 \to e^-e^+$ and $D^0 \to e^-\mu^+$ in $U(1)'$ model (a) and unparticle (b). }
\label{deehh}
\end{figure}

For the decay processes $D^0 \to e^-e^+$ and $D^0 \to e^-\mu^+$,
corresponding Feynman diagrams are shown in Fig.\ref{deehh}.

The corresponding Feynman amplitude with $Z'$ as the mediate
particle is written as
\begin{equation}
\begin{array}{rl}
\mathcal{M}(D^0 \to l_i \bar l_j)=Tr[\bar v(q_2) (\varepsilon_{cu}^L P_L+\varepsilon_{cu}^R P_R)\gamma^\sigma u(q_1)] \frac{1}{m_D^2-m_{Z'}^2}
\bar u(p_1)(\omega_{ij}^L P_L+\omega_{ij}^R P_R)\gamma_\sigma v(p_2)
\end{array}
\end{equation}
where $\omega_{ij}=\omega_{ee}$ for  $D^+_s \to K^+ e^-e^+$ and
$\omega_{ij}=\omega_{e\mu}$ for  $D^+_s \to K^+ e^-\mu^+$. Following
Ref.\cite{Hao:2006nf} we have
\begin{equation}
\begin{array}{rl}
u(q_1) \bar v(q_2) \to f_D (\rlap /p+m_D)\gamma_5 .
\end{array}
\end{equation}
The decay width $\Gamma(D^0 \to e^-e^+)$ is
\begin{equation}
\begin{array}{rl}
\Gamma=&\frac{(\varepsilon_{cu}^L-\varepsilon_{cu}^R)^2(\omega_{ee}^L-\omega_{ee}^R)^2
f_D^2 m_e^2 \sqrt{m_D^2-4m_e^2}}{2\pi (m_{Z'}^2-m_D^2)^2}.
\end{array}
\end{equation}

\subsection{contribution from unparticle}

$D^0 \to e^-e^+$ and $D^0 \to e^-\mu^+$ could also be realized via exchanging a
scalar unparticle, and the corresponding Feynman amplitude is
\begin{equation}
\begin{array}{rl}
\mathcal{M}=Tr[\bar v(q_2) \rlap /p(1-\gamma_5) u(q_1)]
\frac{c^{cu}_s}{\Lambda_\mathcal{U}^{d_\mathcal{U}}}\frac{1}{(m_D^2)^{2-d_\mathcal{U}}}e^{-i(d_\mathcal{U}-2)\pi}
\bar u(p_2)\rlap /p(1-\gamma_5)
v(p_1)\frac{c^{ee}_s}{\Lambda_\mathcal{U}^{d_\mathcal{U}}},
\end{array}
\end{equation}
where $c^{ij}_s=c^{ee}_s$ for $D^0 \to e^-e^+$ and
$c^{ij}_s=c^{e\mu}_s$ for  $D^0 \to e^-\mu^+$.

The decay width $\Gamma(D^0 \to e^-e^+)$ is
\begin{equation}
\begin{array}{rl}
\Gamma=&\frac{(c^{cu}_s c^{ee}_s)^2f_D^2
\sqrt{m_D^2-4m_e^2}me^22^{9-4d_\mathcal{U}}\pi^{4-4d_\mathcal{U}}}
{m_D^{4-4d_\mathcal{U}}\Lambda_\mathcal{U}^{4d_\mathcal{U}}}
\frac{\Gamma^2[1/2+d_\mathcal{U}]} {\sin^2 d_\mathcal{U}\pi
\Gamma^2[d_\mathcal{U}-1]\Gamma^2[2d_\mathcal{U}]}.
\end{array}
\end{equation}

\section{Numerical results}

For $D^+_s \to K^+ e^-e^+$ and $D^+_s \to K^+ e^-\mu^+$ where
a $Z'$ boson is exchanged at s-channel, we follow the authors of
Ref.\cite{Yue:2016mqm,Gao:2010zzg} and set the ranges of
$\varepsilon_{cu}^L$, $\varepsilon_{cu}^R$, $\omega_{ee(\mu)}^L$ and
$\omega_{ee(\mu)}^R$ to $-0.5\sim0.5$ accordingly.

\begin{figure}[H]
\centering
\begin{minipage}[!htbp]{0.8\textwidth}
\centering
\includegraphics[width=0.98\textwidth]{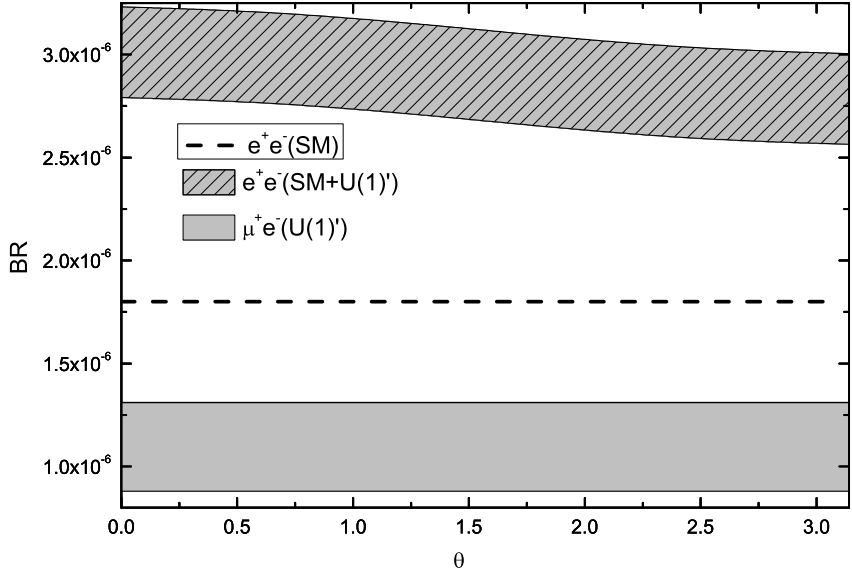}
\caption{The branching ratios of  processes $D^+_s \to K^+ e^-e^+$
and $D^+_s \to K^+ e^-\mu^+$ versus the mixing angle $\theta$
between SM and $U(1)'$  with
$\varepsilon_{cu}^L=\varepsilon_{cu}^R=\omega_{ee}^L=\omega_{e\mu}^L=\omega_{ee}^R=\omega_{e\mu}^R=0.2$
and $m_{Z'}=2000$GeV. The theoretical uncertainty comes from the
form factors. } \label{theta}
\end{minipage}
\end{figure}

We plot the branching ratios of $D^+_s \to K^+ e^+e^-$ and $D^+_s
\to K^+ e^-\mu^+$ versus the mixing angle  between SM $Z$ and
$Z'$ of $U(1)'$ $\theta$ in Fig.\ref{theta}.



\begin{figure}[H]
\centering
\begin{minipage}[!htbp]{0.8\textwidth}
\centering
\includegraphics[width=0.98\textwidth]{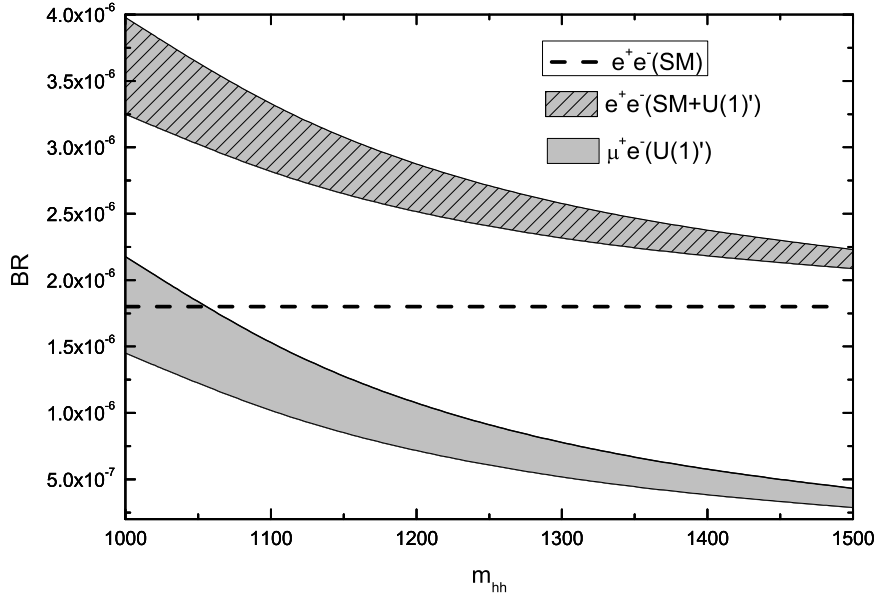}

\caption{The branching ratios of  $D^+_s \to K^+ e^-e^+$ and $D^+_s
\to K^+ e^-\mu^+$ versus the mass of the heavy Higgs boson with
$\rho_{cu}=\rho_{ee}=\rho_{e\mu}=0.15$. The theoretical uncertainty
comes from the form factors.} \label{brhh}
\end{minipage}
\end{figure}

When we calculate the branching ratios of $D^+_s \to K^+ e^-e^+$ and
$D^+_s \to K^+ e^-\mu^+$ via exchanging a heavy Higgs boson, we just
follow Ref.\cite{Liu:2015oaa} and take the range of $\rho^f_{ij}$
within $0.01\sim0.3$, other than adopting the  so-called Cheng¨CSher
ansatz for the couplings $\rho^f_{ij}$ which were done in
Ref.\cite{Cheng:1987rs}. We plot the branching ratios of $D^+_s \to
K^+ e^+e^-$ and $D^+_s \to K^+ e^-\mu^+$ versus the mass of the
heavy Higgs boson in Fig.\ref{brhh}.

\begin{figure}[H]
\centering
\begin{minipage}[!htbp]{0.8\textwidth}
\centering
\includegraphics[width=0.98\textwidth]{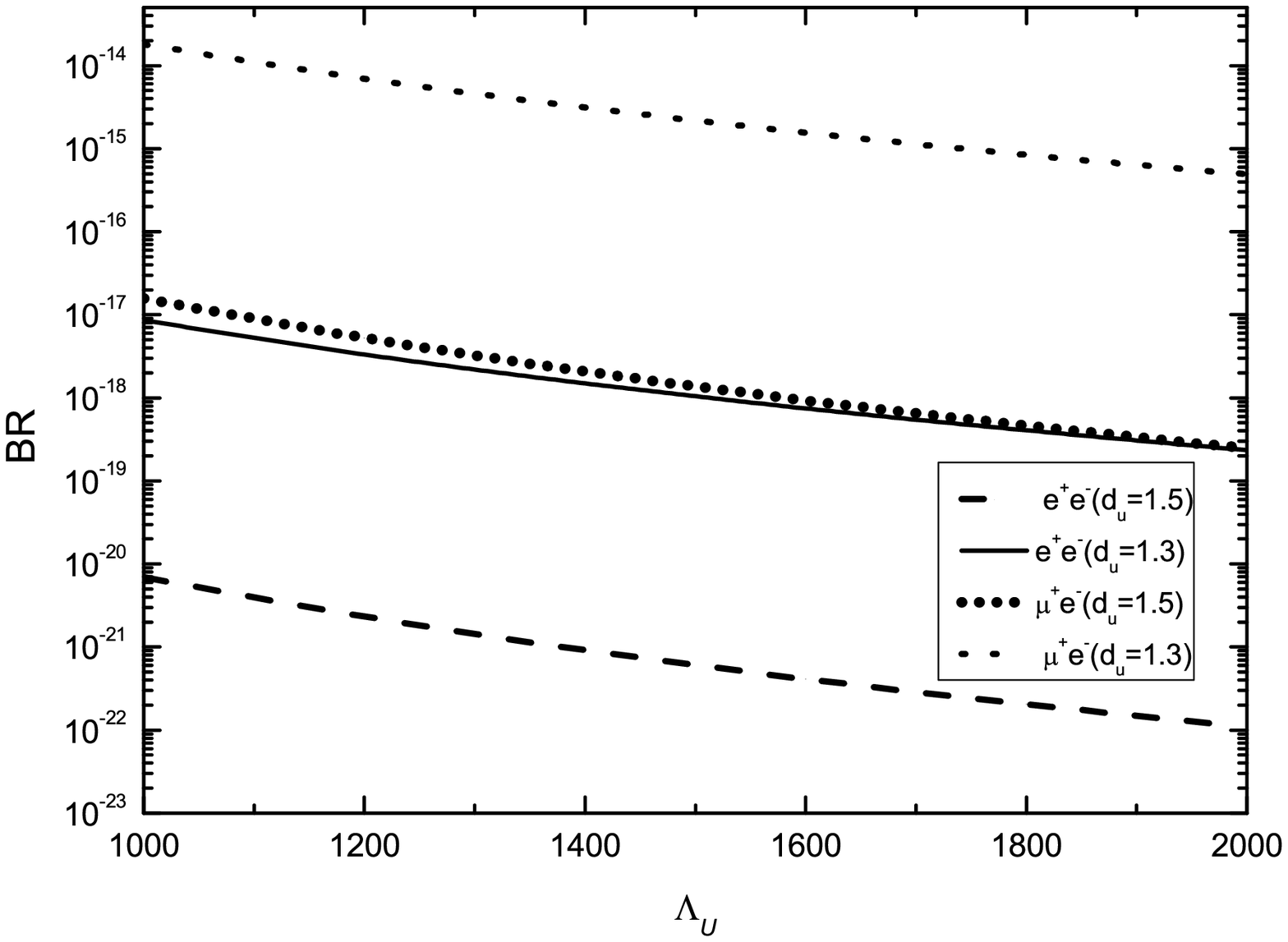}

\caption{The branching ratio of  $D^+_s \to K^+ e^-e^+$ and $D^+_s
\to K^+ e^-\mu^+$ versus the energy scale $\Lambda_\mathcal{U}$ with
$c_S^{cu}=c_S^{e\mu}=0.04$, $c_S^{ee}=0.12$, $d_\mathcal{U}=1.3$ and
$1.5$. } \label{brlbm}
\end{minipage}
\end{figure}

Then, we calculate branching ratios of  $D^+_s \to K^+ e^-e^+$ and
$D^+_s \to K^+ e^-\mu^+$  via exchanging a scalar unparticle.
Following
Refs.\cite{Li:2007by,Jia:2013haa,Cheung:2007zza,Luo:2007bq}, we take
$\Lambda_\mathcal{U}=1$TeV, $1<d_\mathcal{U}<2$ and the range of
$c_S$ to be $0.01\sim0.04$ with the relation
\begin{equation}
\begin{array}{rl}
c_S^{f'f}=
\begin{cases}
c_S     & \text{$f\neq f'$}\\
\kappa c_S & \text{$f=f'$}
\end{cases}
\end{array}
\end{equation}
where $\kappa=3$ \cite{Li:2007by}. Then we plot the branching ratio
of decays  $D^+_s \to K^+ e^-e^+$ and $D^+_s \to K^+ e^-\mu^+$
versus $\Lambda_\mathcal{U}$ with different $d_\mathcal{U}$ in
Fig.\ref{brlbm}.

We list the branching ratios of  $D^+_s \to K^+ e^-e^+$ and $D^+_s
\to K^+ e^-\mu^+$ predicted by various new physics  models (BSM) in
Tab.\ref{tabee} and \ref{tabemu} separately. From those tables we
notice that for the $ U(1)'$ model\cite{Gao:2010zzg} and 2HDM  of
type III\cite{Liu:2015oaa}, the branching ratios can be up to order
of $10^{-6}\sim10^{-7}$.
\begin{table}[htbp]
\begin{center}
\footnotesize
\begin{tabular*}{160mm}{c@{\extracolsep{\fill}}cccc}\hline
          model                   &mass                           & couplings constants  & BR  \\\hline
  $ U(1)'$ \cite{Gao:2010zzg}      &$1000\sim2000$GeV         &$-0.5\sim0.5$      & $10^{-8}\sim10^{-6}$ \\
2HDM type III\cite{Liu:2015oaa}    &  $1000\sim1500$GeV           & $0.05\sim0.3$      &$10^{-8}\sim10^{-6}$ \\
unparticle                   &$1000\sim2000$GeV   & $0.02\sim0.04$           & $10^{-21}\sim10^{-18}$  \\\hline
\end{tabular*}
\caption{ Branch ratios of process $D^+_s \to K^+ e^-e^+$ in different kinds of new physics beyond SM.}
\label{tabee}
\vspace{0mm}
\end{center}
\vspace{0mm}
\end{table}

\begin{table}[htbp]
\begin{center}
\footnotesize
\begin{tabular*}{170mm}{c@{\extracolsep{\fill}}ccccc}\hline
          model                   &mass  & couplings constants   & BR  \\\hline
  $ U(1)'$ \cite{Gao:2010zzg}       &$1000\sim2000$GeV         &$-0.5\sim0.5$        & $10^{-8}\sim10^{-6}$ \\
2HDM type III\cite{Liu:2015oaa}    &  $1000\sim1500$GeV           & $0.05\sim0.3$      &$10^{-8}\sim10^{-6}$ \\
unparticle                         &$1000\sim2000$GeV        & $0.02\sim0.04$        & $10^{-18}\sim10^{-15}$  \\\hline
\end{tabular*}
\caption{ Branching ratios of  $D^+_s \to K^+ e^-\mu^+$ predicted by
various models of new physics beyond SM.} \label{tabemu}
\vspace{0mm}
\end{center}
\vspace{0mm}
\end{table}

We also list  branching ratios of leptonic decays $D^0 \to e^-e^+$
and $D^0 \to e^-\mu^+$ predicted by various models of new physics
beyond SM in Tab.\ref{tabe0}. Since $D^0$ cannot decay to $l_i \bar
l_j$ through a scalar particle, only $Z'$ and unparticle could
contribute to those leptonic decays.

\begin{table}[htbp]
\begin{center}
\footnotesize
\begin{tabular*}{170mm}{c@{\extracolsep{\fill}}ccccc}\hline
decay & &$D^0 \to e^-e^+$   &&\\\hline
          model                   &mass  & couplings constants   & BR  \\\hline
  $ U(1)'$        &$1000\sim2000$GeV         &$-0.5\sim0.5$        & $10^{-13}\sim10^{-10}$ \\
unparticle     &  $1000\sim2000$GeV           & $0.02\sim0.04$      &$10^{-16}\sim10^{-14}$ \\\hline
decay & &$D^0 \to e^-\mu^+$\\\hline
  $ U(1)'$       &$1000\sim2000$GeV         &$-0.5\sim0.5$        & $10^{-9}\sim10^{-7}$ \\
unparticle        &$1000\sim2000$GeV         & $0.02\sim0.04$        & $10^{-11}\sim10^{-9}$  \\\hline
\end{tabular*}
\caption{ Branching ratios of  $D^0 \to e^-e^+$ and $D^0 \to
e^-\mu^+$ predicted by $U(1)'$ and unparticle models.} \label{tabe0}
\vspace{0mm}
\end{center}
\vspace{0mm}
\end{table}

Our numerical results indicate that as the experimental bounds being
taken into account and the corresponding coupling constants in
$U(1)'$ model and 2HDM taking their maximum values, the branching
ratios of  $D^+_s \to K^+ e^-e^+$ and $D^+_s \to K^+ e^-\mu^+$ can
be up to order of $10^{-6}$. Whereas the contribution of the scalar
unparticle to the branching  ratios can only reach an order of
$10^{-18}(10^{-15})$.

\section{Searching for semi-leptonic and leptonic  decays based on the large database of BESIII}

In this section, let us discuss possible constraints and the potential to
observe the aforementioned rare semi-leptonic and leptonic decays of D mesons
based on the large database of BES III. Unlike the hadron colliders, electron-positron
colliders have much lower background which is well understood at present and
helps to reduce contaminations from the measurement circumstance.
Thus controllable and small systematic uncertainties are expected.

The BES III experiment has accumulated large data samples at 3.773 and 4.18 GeV,
which are just above the production thresholds of $D\bar D$ and $D_s^{\star+} D_s^-+c.c.$.
This provides an excellent opportunity to investigate the decays of these charmed mesons.

At these energies, the charmed mesons are produced in pairs. That is to say, if only one
charmed meson is reconstructed in an event, which is defined as a single tag event, there
must exist another charmed meson in the recoiling side. With the selected singly tagged events,
the concerned rare charm decays can be well studied in the recoiling side of the reconstructed charmed meson.

This is named as the double-tag technology, which is firstly employed by the MARK-III Collaboration
and now widely used in the BES III experiments. With this method, the two charmed mesons are both
tagged in one event, one of the charmed mesons is reconstructed through a well measured hadronic channel,
then the other one decays into the concerned signal process.  Benefiting from the extremely clean background,
the systematic uncertainties in double tag measurements can be reduced to a fully controlled level.

In principle, there are two ways to perform the search for rare/forbidden decays.
One is based on the single tag method where one charmed meson is reconstructed for the signal process
while no any constraint is set to the other. This method can provide larger statistics meanwhile a more complex
and higher background might exist as the price to pay. Another way is using the double tag method which presents
a simple and clean
backgrounds but a relatively poorer statistics (see table \ref{tab::technique}). Whether employing the double-tag technique
for studying the relevant processes depends on a balance between reducing background contaminations and expecting higher statistics.

\begin{table*}[htbp]
\begin{center}
\begin{tabular}{lccccccc}
\hline
Method & Statistics (charged/neutral) & Background & Sensitivity \\
\hline
Single Tag Method & $1.7\times10^7$/$2.1\times10^7$ & not good  & Bkg. vs Stat. \\
Double Tag Method & $1.6\times10^6$/$2.8\times10^6$ & clean     & Bkg. vs Stat. \\
\hline
\end{tabular}
\caption{Two methods on searching for rare/forbidden $D$ decays.}
\label{tab::technique}
\end{center}
\end{table*}

In the following, we discuss the statistics of the measurements on the rare decays, which may compose the
factor to restrict the ability of searching for new physics in most cases. For single tag method,
the background analysis is severely mode dependent. Thus, to simplify the estimation, we will focus
our discussion on the result of double tag method.
The BESIII experiment has accumulated huge threshold data samples of about $2.95$ $fb^{-1}$ and $3.15$ $fb^{-1}$
at the cms energies $\sqrt{s}=$ 3.773 and 4.180 GeV, which are about 3.5 times and 5 times more than the previously
accumulated database, respectively.  According to
the published papers of the BESIII experiments, there are more than $1.6\times10^6$ and $2.8\times10^6$ singly tagged
charged and neutral $D\bar D$ mesons, respectively. These modes can be used as the tagging side for
the double-tag method. Namely, because of the advantage of the double-tag method which may remarkably reduce
the background and enhance the confidence level, we suggest to adopt the double-tag method for the analysis on
the rare decay data while employing the well established modes as the tagging side. Then at the recoiling side, one can
look for the expected signal. Omitting some technical details, we know that while adopting this double-tag method, the
experimental sensitivity can reach
about $10^{-6}$ at 90\% Confidence Level (CL)  if assuming zero-signal and zero-background events.
In next 10 years, 4 to 6 times more charm data can be expected, we may have a better chance to detect
such rare decays.

However unfortunately according to our predictions this sensitivity is still below the bound of observing the pure leptonic
rare decays of $D^0$ (no matter lepton-flavor-conserving or lepton-flavor-violating processes).
If the size of BESIII data sample can reach 20 fb$^{-1}$
in next 10 years, the sensitivity would be at $10^{-7}$ level which is almost touching the bottom line of our prediction on the
rate of pure leptonic modes. The analysis is a little more complex at the 4.180 GeV even though the method is similar. The
sensitivities for the rare semi-leptonic decays of $D_s^+$ or $D_s^{\star+}$ mesons can be expected to reach $10^{-5}$ at 90\% CL, however it is
not enough to test our predictions for the rare $D_s^+$ semi-leptonic decays.

If the proposed super $\tau$-charm factory is launched in the near future, we would be able to collect at
least 100 or 1000 times more data since the designed luminosity of STCF will be as high as $1\times10^{35}$ ${\rm cm}^{-2}{\rm s}^{-1}$
which is 100 times of the BEPC II. Then, the sensitivities of searching for the concerned signals in $D$
or $D_s^+$ decays can be greatly improved as $10^{-9 \sim -10}$ or $10^{-7 \sim -8}$ at 90\% CL, respectively, may be expected.
With this improved sensitivities, the rates of $D_s^+\to K^+e^+e^-$ and $D_s^+\to K^+\mu^+e^-$ predicted by
the $U'(1)$ or 2HDM models become measurable. Then, the more challengeable lepton-flavor-violation modes $D^0\to e^-\mu^+$ predicted by
the $U'(1)$ and unparticle models can be possibly tested.

\section{discussion and conclusion}

The rare decays of heavy flavored hadrons which are suppressed or even forbidden in SM can
serve as probe portals for searching new physics BSM. Experimentally measured ``anomalies'' which obviously
deviate from the SM predictions are considered as the candidate signals of BSM, at least provide hints to
BSM for the experiments of high energy colliders, such as LHC. That is the common sense for experimentalists and theorists
of high energy physics. However, how to design a new experiment which might lead to discovery of new physics is an art.
As following the historical experience, beside the blind search in experiments, researchers tend to do measurements according to the
prediction made by theorists based on the available and reasonable models.

FCNC/LFV processes provide a sensitive test for new physics BSM, which compose a
complementary area to high energy collider physics. Definitely,
those processes where SM substantially contributes, do not stand as candidates for seeking
new physics BSM, because the new physics contributions would be drown in the SM background.
Researchers are carefully looking for rare processes where SM contributions are very
suppressed or even forbidden by some rules. The rare semi-leptonic and leptonic decays of
B and D mesons are ideal places because they are caused by FCNC. Especially
the lepton-flavor violation decays which cannot be resulted in by the SM because neutrino masses are too tiny
to make any non-negligible contribution, are the goal which we have interests in.

Recently, most of researches focus on B decays. The reason is obvious, that B mesons are at least three times heavier
than D mesons, so the processes involving B-mesons are closer to new physics scale and moreover, the coupling between b-quark
and top-quark has a large CKM entry. Indeed, there are many research works concerning $B\to K^{(*)}l\bar l$\cite{Ali:1999mm,Altmannshofer:2008dz} and $B^0(B_s)\to
l\bar l$ have emerged\cite{Buchalla:1998ba,Bobeth:2001sq}. On another aspect, several authors have studied the case of D mosons, and drawn constraints
on the free parameters in the proposed models by fitting available data. The model parameters can be compared with those
obtained by fitting the data of B decays. In this work,  based on the large database of the BESIII, we
follow the trend to investigate possibilities
of detecting the rare semi-leptonic and pure leptonic decays of D meson, and  specially we pay more attention to the
analysis of the lepton-flavor-violation processes.

In this work, we calculate the decay rates of $D^+_s \to K^+
e^-e^+$, $D^+_s \to K^+ e^-\mu^+$, $D^0 \to e^-e^+$ and $D^0 \to
e^-\mu^+$ through exchanging a neutral particle in terms of three
BSM new physics models: the extra $U'(1)$, 2HDM of type III and unparticle.
The decay rate of $D^+_s \to K^+ e^-e^+$ receives sizable
contribution from SM whose branching ratio is up to orders of $10^{-6}$.
It is noted that the branching ratio of direct decay process via penguin diagram is small
at order of $10^{-8}$, while the long-distance reaction makes a larger contribution.
Our numerical results show that $U(1)'$ and 2HDM of type
III can make significant contributions to the process $D^+_s \to K^+
e^-e^+$ as long as the model parameters which are obtained by fitting relevant data are adopted,
but the unparticle model cannot make any substantial contribution.
The recent researchers seem to be more tempted to use the extra $U'(1)$ model and we follow
their trends. But here for fixing the model parameters, we deliberately relax the constraint
set by the $D^0-\bar D^0$ mixing as we discussed in the above text. If the constraints were taken into
account, the predicted branching ratio of $D^+_s \to K^+ e^-e^+$ would be reduced by two more orders
as $10^{-8}$ which is much lower than the contribution of the SM long-distance effect. Thus the new physics
contribution would be buried in the SM background. However, as we only consider the constraints on $U'(1)$
parameter taken by fitting the data of $\tau\to 3l$ other than $D^0-\bar D^0$ mixing, the predicted branching ratio
can be large to order of $10^{-6}$, thus the resultant amplitude might interfere with the SM long-distance
contribution.

In future BES III experiment, the
experimental sensitivity can be up to order of $10^{-6}\sim10^{-7}$, thus the data
on $D^+_s \to K^+ e^-e^+$ might tell us some information of
new physics.

Our numerical results show that the $U(1)'$ model and 2HDM of type III could
make an observable branching ratio of $D^+_s \to K^+ e^-\mu^+$ with
the BES III data as its precision
can reach orders of $10^{-6}\sim10^{-7}$.

For processes $D^0 \to e^-e^+$ and $D^0 \to e^-\mu^+$,
the theoretically predicted ranching ratio of decay $D^0 \to
e^-e^+$ is of the order of $10^{-10}$ since its width is proportional to
$m_e^2$, such a small value is hard to be observed.
While for decay $D^0 \to e^-\mu^+$, its branching ratio can be up to
orders of $10^{-7}$, which may be observed in future super
$\tau$-charm factory. Moreover, one can expect to watch  $D^0 \to \mu^-\mu^+$, while
unfortunately,  $D^0 \to \mu^-\tau^+$ is forbidden by the phase space
of final states because $m_{\mu}+m_{\tau}>m_{D^0}$.

According to the presently available new physics models, $U'(1)$, 2HDM and
unparticle model, the data on D-mesons which will be collected in the future 10 years can
marginally detect the new physics contributions to $D^+_s \to K^+ e^-e^+$, $D^+_s \to K^+ e^-\mu^++h.c.$,
$D^0 \to e^-e^+$ and $D^0 \to e^-\mu^++h.c.$ as long as only the constraints set by some experiments are accounted,
but the data of $D^0-\bar D^0$ mixing are relaxed. If the data of $D^0-\bar D^0$ mixing are taken into account
the BES III and even the planned high luminosity $\tau$-charm factory will not be able to ``see'' those rare decays
as predicted by these models. However, it by no means forbids experimental search for these rare decays
in the charm energy regions based on the huge data sample collected by BES and the future $\tau$- charm factory. Blind
experimental search is not affected by the available theoretical prediction because the present BSM
are only possible ones conjectured by theorists, while nature might suggest an alternative scenario. Once such new observation is made,
we would be stunned and explore new models BSM to explain the phenomena, thus our theories would make new
progress, and that is what we expected.

\section*{Acknowledgments}

This work is supported by National Natural Science Foundation with
contract No. 11675082, 11375128,  11405046 and the
Special Grant of the Xuzhou University of Technology No.
XKY2016211.

\end{document}